\documentclass[
nopreprint
]{jasatex}

\usepackage{color}   
\usepackage{graphicx}
\usepackage{dcolumn}
\usepackage{amsmath,amsfonts,amssymb}
\usepackage{bm}

\newcommand{\beq}[1]{  \begin{equation} \label{#1} }  
\newcommand{\eeq}{     \end{equation}} 	
\newcommand{\bal}[1]{  \begin{align} \label{#1} }  
\newcommand{\rf}[1]{(\ref{#1})}
\def\bd#1{\mbox{\boldmath$\displaystyle\mathbf{#1}$} }
\def\tr{\operatorname{tr}} 
\def\dd{\operatorname{d}}

\def\div{\operatorname{div}} 
\def\Div{\operatorname{Div}} 
\def\Nabla{\overline{\nabla}} 
\def\Curl{\operatorname{Curl}} 
\def\axt{\operatorname{axt}} %
\def\curl{\operatorname{curl}}  
\newtheorem{thm}{Theorem}
\newtheorem{lem}{Lemma}
\def\rev{}	   
\begin{document} 
\title{Acoustic metafluids}  

\author{Andrew N. Norris}
  \email{norris@rutgers.edu}
\affiliation{Mechanical and Aerospace Engineering, Rutgers University, Piscataway NJ 08854}

\date{\today}

\begin{abstract}
 
Acoustic metafluids are defined as the class of fluids that allow one domain of fluid to acoustically mimic another, as exemplified by acoustic cloaks.   It is shown that the most general class of 
acoustic metafluids are materials with anisotropic inertia and the elastic properties of what are known as pentamode materials.  
The derivation uses the notion of finite deformation to define the transformation of one region to another.  The main result is found by  considering  energy density in the original and transformed regions.   Properties of acoustic metafluids are discussed, and general conditions are found which ensure that the mapped  fluid has isotropic inertia, which potentially opens up the possibility of achieving broadband cloaking. 

\end{abstract}
\maketitle

\section{Introduction}\label{sec1}

\begin{figure}[ht]  
\begin{center} \includegraphics[width=2.0in]{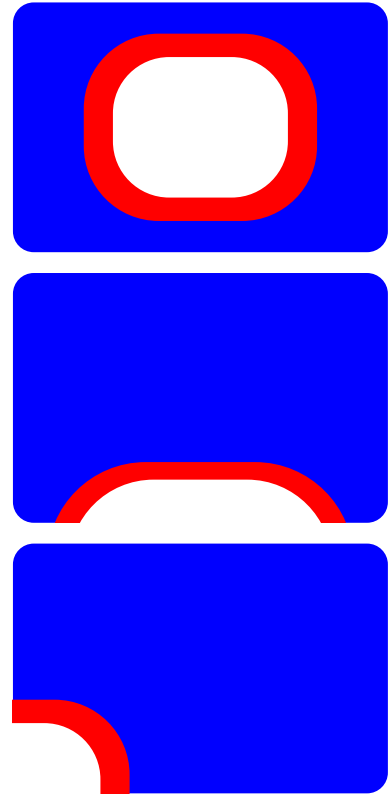} 
\end{center} 
\caption{Three ways to   acoustically hide something: envelope it with a cloak (top);   hide it under a carpet (middle); or hide the object in the corner.  In each case the acoustic metafluid (red) emulates the acoustic properties of the hidden region as if it were filled with the exterior fluid (blue).  }
 \label{fig1}
\end{figure}

\begin{figure}[ht]  
\begin{center} \includegraphics[width=2.8in]{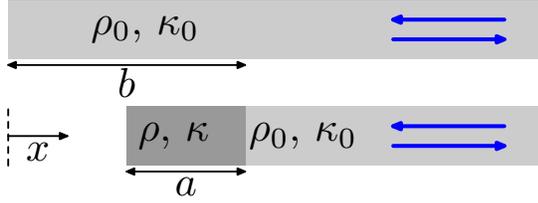} 
\end{center} 
\caption{The top shows waves in a semi-infinite medium $x\ge 0$.  The wave incident from the right reflects from a perfectly reflecting boundary at $x=0$.  The lower figure shows the same medium in $x>b$ with  the region $0\le x< b$ 
replaced by a shorter region of  acoustic metafluid.  Its properties  are such that it produces a perfect  acoustic illusion or ``mirage".}
 \label{fig2}
\end{figure}

Ideal acoustic stealth is provided by the acoustic cloak,  a shell of material that surrounds the object to be rendered acoustically ``invisible".  Stealth can also be achieved by ``hiding under the carpet"\cite{Pendry08} as shown in Fig.~\ref{fig1}.  A simpler situation but one that displays the essence of the acoustic stealth problem is depicted in Fig.~\ref{fig2}.  The   common issue is how to make one region of  fluid acoustically mimic another region of fluid.  The fluids are different as are the domains they occupy; in fact the mimicking region is typically smaller in size, it can be viewed as a compacted version of the original. 

The subject of this paper is not acoustic cloaks, or carpets, or  ways to hide things, but rather the type of material necessary to achieve  stealth.  \rev{We define these materials  as 
\textit{acoustic metafluids}, which as we will see can be considered  fluids with microstructure and properties outside those found in nature.} 
The objective is to derive the general class of acoustic metafluids, and in the process show that there is a closed set which can be mapped from one to another.  Acoustic metafluids are defined as the class of fluids that (a) acoustically mimic another region as in the  examples of Figs. \ref{fig1} and \ref{fig2}, and (b) can themselves be mimicked by another acoustic metafluid in the same sense.  The requirement (b) is important, implying that  there is a closed set of acoustic metafluids.  The set includes as a special case the ``normal" acoustic fluid of uniform density and bulk modulus.  Acoustic metafluids can therefore be used to create stealth devices in a normal fluid.  But, in addition, acoustic metafluids can provide stealth in any type of acoustic metafluid.   The reciprocal nature of these fluids make them a natural generalization of normal acoustic fluids.  

The acoustic cloaks that have been investigated to date fall into two categories in terms of the type of acoustic metafluid proposed as cloaking material.  \rev{Most studies, e.g.  
\cite{Cummer07,Chen07,Cai07,Cummer08,Greenleaf08,Norris08c},  
consider the cloak to comprise fluid with the normal stress-strain relation but anisotropic inertia, 
what we call {\it inertial cloaking}.  Particular realization of inertial cloaks are in principle feasible using layers of isotropic ``normal" fluid \cite{Cheng08,Torrent08,Torrent08b,Chen08a}; the  layers  are introduced in order to achieve an homogenized medium that approximates a  fluid with anisotropic inertia. 
An alternative and  more general approach \cite{Norris08b,Milton06} is to consider anisotropic inertia  combined  with anisotropic elasticity.  The latter is introduced by generalizing the stress strain relation to  include what are known as pentamode\footnote{The name pentamode is based on the defining property that the material supports five easy modes of infinitesimal strain. See \S \ref{sec5} for details.}  elastic materials \cite{Milton95,Milton06,Norris08b}. }   Clearly, the question of how to design and fabricate acoustic metafluids remains open.   The focus of this paper is to first characterize the acoustic metafluids as a general type of material.  In fact, as will be shown, this class of materials contains  broad degrees of freedom, which can significantly aid in future design studies.  

 The paper is organized as follows.  The concept of acoustic metafluids is introduced in \S \ref{sec2} through two  ``acoustic mirage" examples.  The methods used to find the acoustic metafluid in these examples are simple but  not easily generalized.   An alternative and far more powerful approach  is discussed in \S  \ref{sec3}: the transformation method.   This  is based on using the change of variables between the coordinates of the two regions combined with differential relations to identify the metafluid properties of the transformed domain.  Leonhardt and Philbin \cite{Leonhardt08} provide  an instructive  review of the transformation method in the context of optics.  The transformation method does not however define the range of material properties capable of being transformed.  This is the central objective of the paper and it is resolved in \S \ref{sec4} by considering the energy density in the original and transformed domains.   
 Physical   properties of acoustic metafluids are discussed in \S \ref{sec5}, including the unusual property that the top surface is not horizontal when at rest under gravity.   The  subset of acoustic metafluids that have  isotropic inertia is considered in \S \ref{sec6}, and a concluding summary is presented in  \S \ref{sec7}.

\section{Acoustic mirages and simple metafluids}\label{sec2}

The defining property  of an \textit{acoustic metafluid} is its ability to  mimic another acoustic fluid that occupies a different domain.  The simplest type of  acoustic illusion   is what may be called an \textit{acoustic mirage} where an observer hears, for example, a reflection from a distant wall, but in reality the echo originates from a closer boundary.  
Two examples of acoustic mirages are discussed next. 

\subsection{1D mirage}

Consider perhaps the simplest configuration imaginable, a one dimensional semi-infinite medium.  The upper  picture in Fig.~\ref{fig2}  shows the left end of an acoustic half space $x\ge 0$ with uniform density  $\rho_0$ and bulk modulus $\kappa_0$.  The wave speed is $c_0= \sqrt{\kappa_0/\rho_0}$.  
Now replace the region $0\le x <b$ with a shorter section $0< b-a\le x <b$ filled with an acoustic metafluid.
\rev{The acoustic mirage effect requires that an observer in $x>b$ hears a response as if the half space is as shown in the top of Fig.~\ref{fig2}.    This occurs if 
 the  metafluid region is such that:  (i)   no reflection occurs at the interface $x=b$, i.e. 
 the acoustic impedance  in the modified  region is the same as before;  and (ii)  
the round trip travel time of a wave incident from the right is unchanged.  The impedance condition  and   the travel time requirement ensure the amplitude and phase of any  signal is exactly the same as in the original half-space, and hence the mirage is accomplished.}

Let the acoustic metafluid have  material properties $\rho $ and $\kappa $, with speed $c = \sqrt{\kappa /\rho }$.  Conditions (i) and (ii) are satisfied if 
\beq{1}
\rho   c   = \rho_0  c_0,
\qquad
\frac{a}{c} = \frac{b}{c_0}, 
\eeq
respectively, implying the density and bulk modulus in the shorter region are 
\beq{2}
\rho    = \frac{b}{a}\rho_0 ,
\qquad
\kappa   =  \frac{a}{b} \kappa_0 .  
\eeq
In this example the acoustic metafluid is  another acoustic fluid, although with different density and bulk modulus.  
Note that the total mass of the metafluid region is unchanged from the original: 
$\rho_0 b=\rho  a$.

\subsection{2D mirage}

\begin{figure}[ht]      
\begin{center} \includegraphics[width=2.8in]{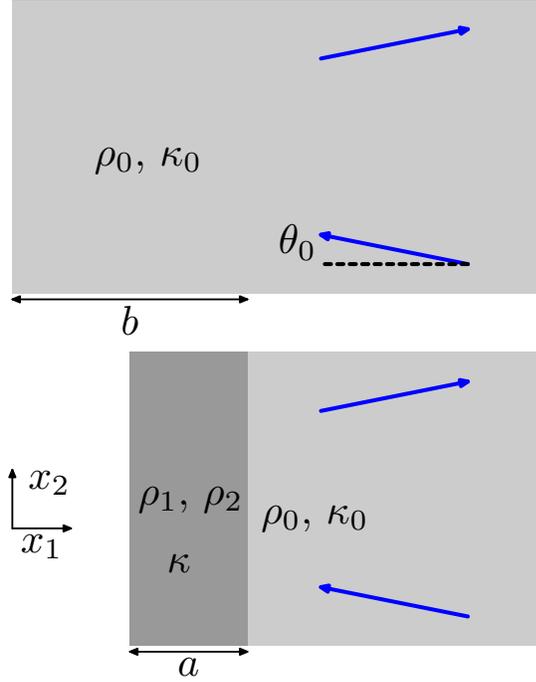} 

\end{center} 
\caption{The same as Fig.~\ref{fig2} but now for oblique incidence.  The same strategy used for the 1D case is no  longer adequate; however a solution may be found if   the metafluid in the layer of thickness $a$ is allowed to have anisotropic inertia defined by the inertias $\rho_1$ and $\rho_2$ in the $x_1$ and $x_2$ directions.}
\label{fig3}
\end{figure}   

Consider the same problem  in 2D under oblique wave incidence, Fig.~\ref{fig3}.  A wave incident at angle $\theta_0$ from the normal has travel time  $b/(c_0 \cos \theta_0 )$ in the original layer. If the shortened region has wave speed $c$, then the modified travel time is  $ a/(c  \cos \theta  )$ where $ \theta  $ is defined by the Snell-Descartes law,   
$\frac1{c }\sin\theta =\frac1{c_0}\sin\theta_0 $.  At the same time, the reflectivity of the modified layer is  $R = (Z-Z_0)/(Z+Z_0)$ where $Z=\rho c /\cos\theta$.  The impedance and travel time conditions are now
\beq{1.1}
\frac{\rho  c}{ \cos\theta }= \frac{\rho_0  c_0 }{\cos\theta_0}   ,
\qquad
\frac{a}{c \cos\theta } = \frac{b}{c \cos\theta_0}.
\eeq
Solving for the modified parameters  implies $\kappa $ is again given by eq. $\rf{2}_2$ but the density is now
\beq{2.0}
\rho   = \frac{b}{a}  \frac{\rho_0}{2\cos^2 \theta_0}
\bigg[ 
1 +   \sqrt{ 1 -  \frac{a^2}{b^2} \sin^2 2\theta_0 }\bigg].
\eeq
The mirage works only for  a single direction of incidence,  $\theta_0$, and is therefore unsatisfactory.    The underlying problem here  is that  three conditions need to be met: Snell-Descartes' law, matched impedances and equal travel times, with only two parameters, $\rho$ and $\kappa$. Some additional degree of freedom is required.

\subsubsection{Anisotropic inertia}
One method to resolve this problem is to introduce the notion of anisotropic mass density, see Fig.~\ref{fig3}.  The density of the metafluid region is no longer a scalar, but becomes a tensor: ${\rho}\rightarrow {\bd \rho}$. 
The equation of motion and the constitutive relation in the metafluid are 
\beq{6}
{\bd \rho}\dot{\bd v} = -\nabla p,
\qquad
\dot{p} = - \kappa \nabla\cdot {\bd v}, 
\eeq
where ${\bd v}$ is particle velocity and $p$ the acoustic pressure. 
Assuming two-dimensional dependence with constant  anisotropic density of the form  
\beq{4}
{\bd \rho} = \begin{bmatrix}
\rho_1 & 0\\
0 & \rho_2
\end{bmatrix} ,
\eeq
and eliminating ${\bd v}$ implies that the pressure satisfies a scalar wave equation
\beq{8}
\ddot{p} - c_1^2 p_{,11} - c_2^2 p_{,22} = 0, 
\eeq
where $c_j = \sqrt{\kappa /\rho_j}$, $j=1,2$.  
Equations \rf{6} and \rf{8}  are discussed in greater detail and generality in \S \ref{sec5}, but for the moment we cite two results necessary for  finding the metafluid in Fig.~\ref{fig3}: 
the phase speed in direction ${\bf n} = 
n_1 {\bd e}_1+n_2 {\bd e}_2$  is $v$, and the associated wave or group velocity vector is ${\bd w}$, where
\beq{008}
v^2 =c_1^2 n_1^2 +c_2^2 n_2^2,
\qquad
{\bd w} = v^{-1} ( c_1^2n_1 {\bd e}_1+ c_2^2 n_2 {\bd e}_2).
\eeq

\subsubsection{Solution of the 2D mirage problem}

The travel time is $a/ {\bd w}\cdot {\bd e}_1$, and the 
 impedance is now $Z = \rho_1 v/\cos\theta$, where, 
 referring to Fig.~\ref{fig3} $n_1 = \cos\theta $ and $n_2 = \sin\theta $.   
Hence, the conditions for zero reflectivity and equal travel times are 
\beq{1.2}
\frac{ \rho_1  v}{ \cos\theta  }= \frac{ \rho_0  c_0 }{\cos\theta_0 }  ,
\qquad
\frac{a v}{c_1^2 \cos\theta }  = \frac{b}{c_0 \cos\theta_0}.
\eeq
Dividing the latter two relations implies $\kappa$ is given by $\rf{2}_2$.  
Snell-Descartes' law,   $
v^{-1}\sin\theta = c_0^{-1}\sin\theta_0$ , 
combined with  eq. $\rf{1.2}_1$ yields
\beq{8.3}
\frac{c_0^2}{v^2} = \frac{ \rho_1^2}{\rho_0^2} 
+\big( 1 - \frac{ \rho_1^2}{\rho_0^2}\big) \sin^2\theta_0 , 
\eeq
while Snell-Descartes' law together with  eq. $\rf{008}_1$ implies 
\beq{8.2}
\frac{c_0^2}{v^2} = \frac{c_0^2 \rho_1}{\kappa } 
+\big( 1 - \frac{  \rho_1}{\rho_2}\big) \sin^2\theta_0 . 
\eeq
Comparison of  \rf{8.3} and \rf{8.2} implies  two identities for 
${ \rho_1}$ and ${ \rho_2}$. 
In summary, the three parameters of the modified region are 
\beq{3.2}
\kappa   =  \frac{a}{b} \kappa_0 ,
\quad
{ \rho_1} =  \frac{b}{a} \rho_0 ,
\quad
{\rho_2} =  \frac{a}{b} \rho_0 .
\eeq
These 
give the desired result: no reflection, and the same travel time for any angle of incidence.  The metafluid layer faithfully mimics the wave properties of the original layer as observed from  exterior vantage points. 

\subsubsection{Anisotropic stiffness}

An alternative solution to the quandary raised by eq. \rf{2.0} is to keep the density isotropic but to relax the standard isotropic  constitutive relation between stress  ${\bd \sigma}$ and strain ${\bd \varepsilon} = [\nabla {\bd u} + (\nabla {\bd u})^t]/2$ 
to allow for material anisotropy.  Thus, the standard relation
${\bd \sigma} = - p {\bd I}$ 
with $p=\kappa \tr {\bd \varepsilon} $ is replaced by the
stress-strain relation for \textit{pentamode materials}
\cite{Milton95} 
\beq{+6}
\bd{ \sigma } = \kappa  \big({\bd Q}: {\bd \varepsilon} \big)\, {\bd Q},
\quad
\div {\bd Q}=0.
\eeq
The physical meaning of the symmetric second order tensor ${\bd Q}$  is discussed later  within the context of a more general constitutive theory. The requirement $\div {\bd Q}=0$ was first noted by Norris \cite{Norris08b} and is discussed in \S \ref{sec5}.  Standard acoustics corresponds to ${\bd Q}= {\bd I}$.

Rewriting $\rf{+6}_1$ as $\bd{ \sigma } =-p {\bd Q}$  and using the divergence free property of ${\bd Q}$, the  equation of motion and the constitutive relation in the metafluid are now
\beq{6-}
\rho\dot{\bd v} = - {\bd Q} \nabla p,
\qquad
\dot{p} = - \kappa {\bd Q}:\nabla {\bd v}. 
\eeq
Eliminating ${\bd v}$ yields the scalar wave equation
\beq{7+}
\ddot{p} - \kappa {\bd Q}: \nabla \big( \rho^{-1} {\bd Q}\nabla p\big) = 0.
\eeq
General properties of this equation have been discussed by Norris \cite{Norris08b} and will be examined later in \S\ref{sec5}.  For the purpose of the problem in Fig.~\ref{fig3} {\bd Q} is assumed constant  of the form 
\beq{4+}
{\bd Q} = \begin{bmatrix}
Q_1 & 0\\
0 & Q_2
\end{bmatrix} ,
\eeq
then it follows that the phase speed and group velocity in direction ${\bd n}$ are 
\beq{8+}
v^2 =C_1^2 n_1^2 +C_2^2 n_2^2,
\quad
{\bd w} = v^{-1} ( C_1^2n_1 {\bd e}_1+ C_2^2 n_2 {\bd e}_2),
\eeq 
where $C_j = Q_j \sqrt{\kappa/\rho}$, $j=1,2$.  
Proceeding as before, using Snell-Descartes' law and the conditions of equal travel time and matched impedance, yields  
\beq{2+}
Q_1^2  =  \frac{a}{b} \frac{\kappa_0}{\kappa}, 
\quad
Q_2^2 =     \frac{b}{a}\frac{\rho}{\rho_0} ,
\quad
Q_1Q_2 =   \frac{b}{a}.  \nonumber
\eeq
Since the important physical quantity  is the product of 
$\kappa$ with ${\bd Q}\otimes {\bd Q}$, any one of 
 the three parameters $\kappa$, $Q_1$ and $Q_2$ may be 
 independently selected.  
\rev{A natural choice  is to impose $\div{\bd Q}=0$ at the interface, which means that the ``flux'' vector ${\bd Q}{\bd n}$ is continuous, where ${\bd n}$ is the interface normal. In this case ${\bd n} = {\bd e}_1$ 
 so that $Q_1=1$ and 
 }
\beq{4-}
\rho    = \frac{b}{a}\rho_0 ,
\quad
\kappa   =  \frac{a}{b} \kappa_0 ,
\quad
{\bd Q} = \begin{bmatrix}
1 & 0\\
0 & \frac{b}{a}
\end{bmatrix} .
\eeq

Comparing the alternative metafluids defined by eqs. \rf{3.2} and \rf{4-} note that in each case   the density and the stiffness associated with the normal ${\bd e}_1$ direction both equal their 1D values given by  eq. \rf{2}.   The first metafluid of eq. \rf{3.2} has a smaller inertia in the transverse direction ${\bd e}_2$ $(\rho_2 <\rho_1)$.  The second metafluid defined by eq. \rf{4-} has increased stiffness in the transverse direction $(Q_2 >Q_1)$.   The net effect in each case is an increased phase speed in the transverse direction as compared with the normal direction $(c_2 >c_1,\,  C_2 >C_1)$.  

The  1D and 2D mirage examples  illustrate the general idea of acoustic metafluids as   fluids that  replicate the wave properties of a transformed region.  However, the methods used to find the metafluids are not easily  generalized to arbitrary regions.  How does one find the metafluid that can, for instance, mimic a full spherical region by a smaller shell?   This is the cloaking problem.  The key to the generalized procedure are the related notions of transformation and  finite deformation, which are introduced next. 

\section{The transformation method}\label{sec3}

\subsection{Preliminaries}
Let $\Omega$ and $\omega$ denote the original and the deformed domains (the regions $0\le X <b$ and $b-a\le x <b$ in the examples of \S \ref{sec2}).  The coordinates in each configuration are
${\bd X}$ and ${\bd x}$, respectively;  the divergence operators  are
$\Div$ and $\div$, and the gradient (nabla) operators  are  $\Nabla$ and 
$\nabla$.  Upper and lower case indices  indicate components, $X_I$, $x_i$ and 
the component form of $\div {\bf A}$ is $\partial A_i /\partial x_i = A_{i,i}$ or 
 $\partial A_{iJ} /\partial x_i =A_{iJ,i}$ when  ${\bf A}$ is a vector and a second order tensor-like quantity, respectively, and repeated indices imply summation (case sensitive!).   Similarly  $\Div {\bf B} = B_{Ij,I}$. 

The finite deformation or transformation is defined by the mapping  
$\Omega \rightarrow  \omega$ according to    ${\bd x} = {\bd \chi}({\bd X})$.
In the terminology of finite elasticity  ${\bd X}$ describes a particle position  in the Lagrangian or undeformed configuration, and ${\bd x}$ is particle location in the Eulerian or   deformed  physical state.  The transformation  ${\bd \chi}$ is assumed to be one-to-one and invertible\footnote{The bijective property of the mapping  does not extend to acoustic cloaks, where there is a single point in $\Omega$  mapped to a surface in $\omega$, see Norris \cite{Norris08b} for details.}.   The deformation gradient is defined 
${\bd F} = \Nabla {\bd x}$  with inverse ${\bd F}^{-1} = \nabla {\bd X}$, 
or in component form $F_{iI}  = \partial x_i /\partial X_I$, $F_{Ii}^{-1}  = \partial X_I /\partial x_i$.    
The Jacobian of the finite deformation is $
\Lambda = \det {\bd F} = |{\bd F}|$,  or in terms of volume elements in the two configurations, $\Lambda = \dd v/\dd V$. The  polar decomposition implies ${\bd F} = {\bd V} {\bd R} $, where the rotation ${\bd R}$ is proper  orthogonal ($ {\bd R}{\bd R}^t = {\bd R}^t {\bd R} = {\bd I}$, $\det {\bd R}= 1$) and the left stretch tensor ${\bd V}\in$Sym$^+$ is the  positive definite solution of ${\bd V}^2 = {\bd F}{\bd F}^t$.
Note for later use the kinematic identities
 \cite{Ogden84}
\beq{+2}
\div \big( \Lambda^{-1} {\bd F}\big) = 0, 
\qquad
  \Div \big( \Lambda  {\bd F}^{-1}\big) = 0,  
\eeq  
and the expression for  the Laplacian in ${\bd X}$ in terms of derivatives in ${\bd x}$, i.e., the chain rule \cite{Norris08b}
\beq{900}
\Nabla^2 f= \Lambda  \div \big( \Lambda^{-1} {\bd V}^2 \nabla f\big). 
\eeq

\subsection{The transformation method} 

The undeformed domain $\Omega$ is of arbitrary shape and comprises 
a homogeneous acoustic fluid with 
density $\rho_0$ and bulk modulus $\kappa_0$. The goal is  to mimic the scalar wave equation in $\Omega$,
\beq{91-}
\ddot{p} - 
(\kappa_0/\rho_0 ) \Nabla^2 p  = 0, \qquad
  {\bd X} \in  \Omega, 
\eeq 
by the wave equation of a metafluid occupying the deformed region $\omega$.  
The basic result \cite{Greenleaf07,Greenleaf08,Norris08b} is that eq. \rf{91-} is exactly replicated in $\omega$ by the  equation   
\beq{933}
\ddot{p} - 
\kappa \div \big( {\bd \rho}^{-1}\nabla p\big)  = 0 , \qquad
  {\bd x} \in  \omega, 
\eeq
where the bulk modulus and inertia tensor are  
\beq{+4}
\kappa = \kappa_0 \Lambda, \qquad
   {\bd \rho} = \rho_0 \Lambda  {\bd V}^{-2}.  
\eeq

The equivalence of eq. \rf{91-} with eqs. \rf{933} and \rf{+4} is evident from the differential equality \rf{900}.  The  idea is to use the change of variables from ${\bd X}$ to ${\bd x}$ to identify the metafluid properties.   Equation \rf{+4} describes a metafluid with anisotropic inertia and isotropic elasticity.  It can be used for modeling acoustic cloaks but has the unavoidable feature that the total \rev{effective}  mass of the cloak is infinite.  This problem, discussed by Norris  \cite{Norris08b}, arises from the singular nature of the finite deformation in a cloak which makes   $\Lambda {\bd V}^{-2}$ non-integrable.   \rev{This type of fluid, which could be called an inertial fluid, appears to be the main candidate considered for acoustic cloaking to date. 
  The major exception is  Milton et al.\cite{Milton06} who considered fluids with properties  of pentamode materials, although as we will discuss in Section \ref{sec4}, their findings are of limited use for acoustic cloaking. 
}

\subsubsection{Pentamode materials}
Norris \cite{Norris08b}  showed that eq. \rf{900} is a special case of a more general identity:
\beq{901}
\Nabla^2 p= \Lambda  {\bd Q}:\nabla \big( \Lambda^{-1} {\bd Q}^{-1}{\bd V}^2 \nabla p\big), 
\eeq
where ${\bd Q}$ is \textit{any} symmetric, invertible and  divergence free $(\div {\bd Q} =0)$ second order tensor. 
The increased degrees of freedom afforded by the arbitrary nature of ${\bd Q}$ means that \rf{91-} is equivalent to the generalized scalar wave equation in $\omega$,
\beq{904}
\ddot{p} - 
\kappa  {\bd Q}:\nabla \big( {\bd \rho}^{-1}{\bd Q} \nabla p\big)  = 0 , \qquad
  {\bd x} \in  \omega, 
\eeq
where the modulus $\kappa$ and the  inertia follow from a comparison of eqs. \rf{91-},  \rf{901} and \rf{904} as 
\beq{+5}
\kappa = \kappa_0 \Lambda, \qquad
   {\bd \rho} = \rho_0 \Lambda  {\bd Q}{\bd V}^{-2}{\bd Q}.  
\eeq
As will become apparent later, these metafluid parameters describe a pentamode material with anisotropic inertia.    For the moment we return to the acoustic mirages in light of the general transformation method.

\subsection{Mirages revisited}

The 2D mirage problem corresponds to  the following finite 
deformation 
$x_1 = b-a + ab^{-1} X_1 $, 
$x_2 = X_2$ for $0\le X_1 < b $, $-\infty < X_2 < \infty $.
The deformation gradient is  
\beq{4.1}
 {\bd F} = \begin{bmatrix}
 \frac{a}{b} & 0 \\
0 & 1 \end{bmatrix},
\eeq
implying ${\bd R} = {\bd I}$,${\bd V} = {\bd F}$ and  $\Lambda =  a/b$.  Equation \rf{+4} therefore implies 
  \beq{4.2}
\kappa  =  \frac{a}{b} \kappa_0 , 
\qquad
 {\bd \rho} = \begin{bmatrix}
 \frac{b}{a}\rho_0
 & 0 \\
0 & \frac{a}{b} \rho_0 \end{bmatrix}.
\eeq
These agree with the parameters found in \S \ref{sec2}, eq. \rf{3.2}. 

Using the more general formulation of \rf{904} and \rf{+5} with the arbitrary tensor chosen as ${\bd Q} = \Lambda^{-1}{\bd V}$ yields the metafluid described by eq. \rf{4-}.   It is interesting to note that although ${\bd \rho}$ of eq. \rf{+5} is, in general, anisotropic, it can be made isotropic in this instance by any ${\bd Q}$ proportional to ${\bd V}$.   Keeping in mind the requirement seen above that $Q_1=0$, we consider
as a second example of \rf{904} the case   
${\bd Q} = \Lambda^{-1}{\bd V}$.  The mirage can then be achieved with material properties
  \beq{4.3}
\rho = \frac{b}{a}\rho_0 , 
\quad
\kappa  =  \frac{a}{b} \kappa_0 
 ,\quad
 {\bd Q}= 
 \begin{bmatrix}
 1  
 & 0 \\
0 &\frac{b}{a} \end{bmatrix}.
\eeq
This again corresponds to a pentamode material with isotropic inertia, equal to that of \rf{4-}.   These two examples illustrate the power associated with the arbitrary nature of the divergence free tensor $\bd Q$.   There appears to be a multiplicative degree of freedom associated with $\bd Q$ that is absent using anisotropic inertia.   As will be evident  later, this degree of freedom is related to a \textit{gauge transformation}.

\section{The most general type of acoustic metafluid}\label{sec4}

\subsection{Summary of main result}

 In order to make it easier for the reader to assimilate, the  paper's main result is first presented in the form of a theorem.   In the following ${\bd Q}_0$ and ${\bd Q}$ are arbitrary 
 symmetric, invertible and divergence free second order tensors.   

\begin{thm}\label{thm1}
\rev{
The  kinetic and strain energy densities in $\Omega $, of the form 
\beq{t1}
T_0 =\dot{\bd U}^t {\bd \rho}_0  \dot{\bd U},
\qquad
W_0= \kappa_0 \big(  {\bd Q}_0:\Nabla {\bd U}\big)^2 , 
\eeq  
respectively, are equivalent to  the current energy densities in $\omega$:
\beq{012}
T= \dot{\bd u}^t {\bd \rho}  \dot{\bd u},
\qquad
W = \kappa \big(  {\bd Q}:\nabla {\bd u}\big)^2 , 
\eeq  
where 
\begin{subequations}\label{91}
\bal{91a}
\kappa &= \Lambda \kappa_0 ,
\\
{\bd \rho} &= \Lambda {\bd Q}{\bd F}^{-t}{\bd Q}_0^{-1} {\bd \rho}_0
{\bd Q}_0^{-1} {\bd F}^{-1}
{\bd Q}.\label{91b}
\end{align}
\end{subequations}
}
\end{thm}
Discussion of the implications are given following the proof. 

\subsection{Gauge transformation}
\rev{
The   energy functions per unit volume in the undeformed  configuration, $ T_0$ and $W_0$, depend upon  the infinitesimal  displacement ${\bd U} $ in that configuration.  The kinetic energy is defined by the density $ {\bd \rho}_0$, while the 
 strain energy is 
$W_0 = C_{0IJKL}U_{J,I}U_{L,K} $,
where  $C_{0IJKL}$  are elements of the stiffness.  The density and stiffness possess the symmetries
$
\rho_{0IJ} = \rho_{0JI}$,
$
C_{0IJKL}= C_{0KLIJ}$, 
$
C_{0IJKL}= C_{0JIKL}$.
The total energy  is $E_0=T_0+W_0$ and 
the total energy $E=T+W$ per unit deformed volume is, using $E_0 \dd V = E \dd v$, simply
$E = \Lambda^{-1}E_0$.  
}

Our objective is to find a general class of material parameters 
$\{  {\bd \rho}_0, \, {\bd C}_0 \}$ that maintain the structure of the energy under a general transformation $\chi$.  Structure here means that the energy  remains  quadratic in  velocity  and  strain.  
In order to achieve the most general form for the transformed energy
  introduce a \textit{gauge transformation} for the displacement.  Let
\beq{05}
{\bd U} = {\bd G}^t {\bd u} ,
\eeq
or $U_I = u_i G_{iI}$ in components, 
where ${\bd G}$ is independent of time but can be spatially varying. Thus, using the chain rule  $U_{J,I } = U_{J,i}F_{iI}$, yields 
$E = T+W$,
where the kinetic and strain energy densities are
\begin{subequations}
\bal{07}
T &= \dot{\bd u}^t  {\bd \rho}  \dot{\bd u}, 
\\
W &= 
\Lambda^{-1}
C_{IJKL} F_{iI}F_{kK} (G_{jJ}u_{j})_{,i}(G_{lL}u_{l})_{,k}  \label{07b}
,
\end{align}
\end{subequations}
and the transformed inertia  is
\beq{08}
{\bd \rho} =   \Lambda^{-1} {\bd G}{\bd \rho}_0{\bd G}^t.
 \eeq
 The kinetic energy  has the required structure, quadratic in the velocity; the strain energy  however is not in the desired form.  The objective is to obtain 
 a strain energy of standard  form 
 \beq{09}
 W = C_{ijkl} u_{j,i} u_{l,k}, 
 \eeq
 where ${\bd C}$ has the usual symmetries: 
$ C_{ijkl}= C_{klij}$, $ C_{ijkl}= C_{jikl}$.

The  question of how to  convert $W$ of \rf{07b} into the form of \rf{09} for generally anisotropic elasticity stiffness ${\bd C}_0$ will be discussed in a separate paper.  The  objective here is to find the largest  class of metafluids that includes all those previously found.  

\subsection{Pentamode to pentamode}
In order to proceed assume that the initial stiffness tensor is of pentamode form \cite{Milton95} 
\beq{10}
 {\bd C}_0 = \kappa_0 {\bd Q}_0\otimes {\bd Q}_0  , 
 \qquad \Div{\bd Q}_0=0, 
\eeq
that is, $ C_{0IJKL}= \kappa_0 Q_{0IJ}Q_{0KL}$ 
where ${\bd Q}_0^t = {\bd Q}_0$ is a positive definite symmetric second order tensor.  The  tensor ${\bd Q}_0$ is necessarily  divergence free
\cite{Norris08b}.

The strain energy density in the physical  space after the 
general deformation and gauge transformation is now 
\beq{12}
W = \kappa_0 \Lambda^{-1} \big[  Q_{0IJ} (u_j G_{jJ})_{,I}\big]^2 .
\eeq  
Consider 
\beq{13}
 Q_{0IJ} (u_j G_{jJ})_{,I} =  Q_{0IJ} G_{jJ} u_{j,I}
 +  Q_{0IJ} G_{jJ,I} u_{j }.
 \eeq
In order to   achieve the quadratic structure of  \rf{09} the final term in \rf{13} must vanish.  Since ${\bd u}$ is considered arbitrary this in turn implies that $Q_{0IJ} G_{jJ,I}$ must vanish for all $j$. 
With no loss in generality let
\beq{15}
{\bd G}^t = {\bd Q}_0^{-1} {\bd P},
\eeq
or $G_{jJ}= Q_{0JM}^{-1}P_{Mj}$ in components. 
Then using the identity for the derivative of a second order tensor, 
$({\bd A}^{-1})_{,\alpha} = - {\bd A}^{-1}{\bd A}_{,\alpha} {\bd A}^{-1}$ where $\alpha$ can be any parameter, gives
\bal{16}
0 & = Q_{0IJ} G_{jJ,I} 
\nonumber \\ & =
- Q_{0IJ}Q_{0JK}^{-1} Q_{0KL,I}Q_{0LM}^{-1}P_{MJ} + P_{Ij,I}
\nonumber \\
& = -  Q_{0IL,I}Q_{0LM}^{-1}P_{Mj} + P_{Ij,I}
\nonumber \\
& = P_{Ij,I}, 
\end{align}
where the important property $\rf{10}_2$ has been used.  
Hence,  $\Div{\bd P}=0$. 
Then  using \rf{13}   yields
\bal{18}
 Q_{0IJ} (u_j G_{jJ})_{,I} &=  P_{Ij} u_{j,I}
\nonumber \\
& = P_{Ij} F_{iI}u_{j,i}
\nonumber \\
& =\Lambda  Q_{ij}u_{j,i},
\end{align}
 where the 
 tensor ${\bd Q}$ is defined by 
 \beq{19}
 {\bd P} = \Lambda {\bd F}^{-1}{\bd Q}
 \qquad\Leftrightarrow 
 \qquad
 {\bd Q} = \Lambda^{-1} {\bd F}{\bd P}.
 \eeq
 The condition $\Div{\bd P}=0$
 becomes, using $\rf{+2}_2$, 
 \beq{21}
 P_{Ij,I} = \big( \Lambda F_{Ii}^{-1}Q_{ij}\big)_{,I}
 = \Lambda F_{Ii}^{-1}Q_{ij,I} 
 = \Lambda Q_{ij,i} , 
 \eeq
 implying 
 \beq{22}
\div{\bd Q}=0.
\eeq

It has therefore been demonstrated  that if the gauge transformation is 
defined by 
\beq{225}
{\bd G}^t = \Lambda {\bd Q}_0^{-1}{\bd F}^{-1}{\bd Q},
\eeq
 where ${\bd Q}$ is divergence free in physical coordinates, then the strain energy \rf{12} is 
$
W = \kappa_0 \Lambda \big(  Q_{ij} u_{j,i}\big)^2 $.
This is of the desired form, eq. \rf{09}, with $ C_{ijkl}= \kappa Q_{ij}Q_{kl}$, hence completing the proof of Theorem \ref{thm1}.

\subsection{Discussion}

\subsubsection{Equivalence of physical quantities}

Theorem \ref{thm1} states that  the pentamode material  defined by stiffness $\kappa_0$ with   stress-like tensor  ${\bd Q}_0$ and anisotropic inertia ${\bd \rho}_0$ is converted into another pentamode material with anisotropic inertia.   
The properties of the new metafluid are defined by  the original metafluid and the  deformation-gauge pair
$\{ {\bd F}, {\bd G}\}$ where ${\bd F}$ is arbitrary and possibly inhomogeneous, and 
$ {\bd G}$ is given by eq. \rf{225} 
with ${\bd Q}$ symmetric, positive definite and divergence free but otherwise completely arbitrary. 
The  special case of a fluid with isotropic stiffness but anisotropic inertia,  eq. \rf{+4},  is recovered from Theorem \ref{thm1} when the starting medium is a standard acoustic fluid and ${\bd Q}$ is taken to be ${\bd I}$.  

It is instructive to  examine  how  physical quantities transform: we consider displacement, momentum and pseudopressure. 
Eliminating ${\bd G}$, it is possible to express the new displacement vector in terms of the original, 
\beq{248}
   {\bd u}= \Lambda^{-1}{\bd Q}^{-1}{\bd F}{\bd Q}_0 {\bd U} .
\eeq
Physically, this means that as the metafluid in $\omega$ acoustically replicates that in $\Omega$, particle  motion in the latter is converted into the mimicked motion defined by \rf{248}.  

Define the momentum vectors in the two configurations, 
\beq{32}
 {\bd m}_0=  {\bd \rho}_0 \dot{\bd U} , 
 \qquad
  {\bd m} =  {\bd \rho}  \dot{\bd u} .
\eeq 
Equations \rf{08} and \rf{225} imply  that they are related by 
\beq{33}
  {\bd m} =    {\bd Q}{\bd F}^{-t}{\bd Q}_0^{-1}{\bd m}_0 .
\eeq
The transformation of momentum is similar to that  for displacement, eq. \rf{248}, but with the inverse tensor, i.e.  ${\bd u}= {\bd G}^{-t} {\bd U} $ while 
$ {\bd m} = \Lambda^{-1}{\bd G}{\bd m}_0$.

Stress in the two configurations is defined by Hooke's law in each:  
\beq{-1}
{\bd \sigma}_0 = {\bd C}_0:\Nabla{\bd U},
\qquad 
{\bd \sigma} = {\bd C}:\nabla{\bd u},
\eeq
where ${\bd C}_0$ and ${\bd C}$  are fourth order elasticity tensors, 
\beq{24}
 {\bd C}_0  = \kappa_0  {\bd Q}_0\otimes {\bd Q}_0  ,
 \qquad 
 {\bd C}  = \kappa  {\bd Q}\otimes {\bd Q}, 
\eeq
that is, $C_{ijkl} = \kappa Q_{ij}Q_{kl}$, etc. 
Using \rf{10}, \rf{18} and \rf{24}, yields
\beq{0-4}
{\bd \sigma}_0 = - p {\bd Q}_0,
\qquad
{\bd \sigma} = - p {\bd Q},
\eeq
where $p$ is the same in each configuration, 
\beq{31}
p = - \kappa {\bd Q}: \nabla{\bd u} =    - \kappa_0 {\bd Q}_0: \Nabla{\bd U} .
\eeq 
The quantity $p$ is similar to pressure, and can be exactly identified as such when ${\bd Q}$ is diagonal, but it is not pressure in the usual meaning.  For this reason it is called the  pseudopressure.   It is interesting to compare the equal values of $p$ in $\Omega $ and $\omega$ with the  more complicated relations \rf{248} and \rf{33} for the displacement and momentum. 

\rev{
\subsubsection{Equations of motion}
The equations of motion can be derived as the Euler-Lagrange equation for the Lagrangian $T-W$. A succinct form is as follows, in terms of the   the momentum density ${\bd m}$ and the  stress tensor ${\bd \sigma}$: 
\beq{41}
\dot{\bd m} = \nabla {\bd \sigma}. 
\eeq
 The constitutive relation may be expressed as an  equation for the pseudopressure $p$ \cite{Norris08b}, 
\beq{42}
\ddot{p} = - \kappa{\bd Q}:\nabla \ddot{\bd u},    
\eeq
while eqs. \rf{31}, \rf{32} and \rf{41} imply that the acceleration is  
 \beq{43}
\ddot{\bd u} = - {\bd \rho}^{-1}{\bd Q}\nabla p.    
\eeq
Eliminating $p$ between the last two equations implies that the displacement satisfies
\beq{+42}
 \kappa{\bd Q}{\bd Q}:\nabla \ddot{\bd u} -    {\bd \rho}\ddot{\bd u} = 0.
\eeq
}
\rev{This is, as expected, the elastodynamic equation for a pentamode material with anisotropic inertia.  Alternatively,  
eliminating $\ddot{\bd u}$ between eqs. \rf{42} and \rf{43}  yields a scalar wave equation for the pseudopressure, 
\beq{44}
\ddot{p} -
\kappa{\bd Q}:\nabla \big( {\bd \rho}^{-1}{\bd Q}\nabla p\big)   = 0.    
\eeq
}
\rev{This clearly reduces to the standard acoustic wave equation when ${\bd Q} = {\bd I}$ and  ${\bd \rho}$ is isotropic. 
}

\rev{
\subsubsection{Relation to the findings of Milton et al.\cite{Milton06}}
}
The present findings appear to contradict those of 
Milton et al.\cite{Milton06} who found the negative result that it is not in general possible to find a metafluid that replicates a standard acoustic medium after  arbitrary finite deformation. 
However, their result is based on the  assumption that  ${\bd G} \equiv {\bd F}$
 (their eq. (2.2)).
Equation \rf{225} implies that ${\bd Q}$ must then be  
\beq{226}
{\bd Q} = \Lambda^{-1} {\bd F}{\bd Q}_0 {\bd F}^t . 
\eeq
Using $\rf{+2}_1$ and $\rf{10}_2$ yields  $
Q_{ij,i} = \Lambda^{-1} Q_{0IJ} F_{jJ,I}$. 
Hence, this particular ${\bd Q}$ can only satisfy the requirement \rf{22} that   
$\div {\bd Q}=0$ if  
\beq{27}
Q_{0IJ}  \frac{\partial^2 x_i}{\partial X_I \partial X_J}= 0. 
\eeq
Milton et al.\cite{Milton06} considered   ${\bd Q}_0$ isotropic (diagonal), in which case \rf{27} means that the only permissible finite deformations are harmonic, i.e. those for which $\Nabla^2 {\bd x} = 0$.
In short, the  negative findings of Milton et al.~\cite{Milton06} are a consequence of constraining the gauge to  ${\bd G} \equiv {\bd F}$, which in turn severely restricts the realizability of  metafluids except under very limited types of transformation deformation.   The main difference in the present analysis is the  inclusion of the general gauge transformation which enables us to find metafluids under arbitrary deformation. 

\section{Properties of acoustic metafluids}\label{sec5}

The primary result of the paper, summarized in Theorem \ref{thm1}, states that 
the class of acoustic metafluids is defined by the most general type of  pentamode material with  elastic stiffness $\kappa {\bd Q}\otimes {\bd Q}$ where $\div {\bd Q}=0$,  and anisotropic inertia ${\bd \rho}$.  We now examine some of the unusual physical properties, dynamic and static,  to be expected in acoustic metafluids.    Some of the dynamic properties have been discussed by Norris \cite{Norris08b}, but apart from Milton and Cherkaev \cite{Milton95} no discussion of static   effects has been given.

\subsection{Dynamic  properties: plane waves}

Consider plane wave solutions for  displacement of the form 
$
{\bd u}({\bd x},t) = {\bd q} e^{ik( {\bd n}\cdot {\bd x} - v t)}$, 
for $|{\bd n}|=1$ and constant ${\bd q}$, $k$ and $v$, and uniform metafluid properties.  
Non trivial solutions    satisfying the equations of motion eq. \rf{+42} require   
\beq{-125}
\big(  \kappa\,  ({\bd Q}{\bd n})\otimes ({\bd Q}{\bd n}) - {\bd \rho} v^2 \big) {\bd q}= 0.  
\eeq
The acoustical or Christoffel \cite{Musgrave}  tensor $\kappa  ({\bd Q}{\bd n})\otimes ({\bd Q}{\bd n})$ is rank one  and it follows that  of the three possible solutions for $v^2$, only one is not zero, the quasi-longitudinal solution
\beq{-104}
v^2 =  \kappa\,  {\bd n}\cdot  {\bd Q} {\bd \rho}^{-1}{\bd Q}{\bd n} , 
\qquad
{\bd q} = {\bd \rho}^{-1}{\bd Q}{\bd n} . 
\eeq
The slowness surface is an ellipsoid. 
 The energy flux velocity \cite{Musgrave}  (or wave velocity or group velocity ) is 
\beq{37}
{\bd w} = v^{-1} \kappa\,    {\bd Q} {\bd \rho}^{-1}{\bd Q}{\bd n}. 
\eeq
${\bd w}$ is in the direction ${\bd Q}{\bd q}$, and satisfies  
${\bd w}\cdot {\bd n} = v$, a well known relation for generally anisotropic solids with isotropic density. 

\subsection{Static properties}

\subsubsection{Five easy modes}
The static properties of acoustic metafluids are just as interesting, if not more so. 
Hooke's law $\rf{0-4}_2$  is 
\beq{0-1}
{\bd \sigma} = {\bd C}{\bd \varepsilon},
\eeq
where ${\bd \varepsilon}={\bd \varepsilon}^t$ is  strain and the stiffness   $ {\bd C} $ is 
defined by eq. $\rf{-1}_2$.    The 
strain energy is 
$
 W = \kappa ( {\bd Q}: {\bd \varepsilon})^2$.
Note that ${\bd C}$ is not invertible in the usual sense of fourth order elasticity tensors. If considered as a $6\times 6$ matrix mapping strain to stress then the stiffness is rank one: 
it has only one non-zero eigenvalue.  This means that there are five independent strains each of which will produce zero stress and zero strain energy, hence the name \textit{pentamode}\cite{Milton95}.   
   The five  ``easy" strains are easily identified in terms of the principal directions and eigenvalues of  ${\bd Q}$.  Let 
 \beq{046}
 {\bd Q} = \lambda_1 {\bd q}_1{\bd q}_1 +\lambda_2 {\bd q}_2{\bd q}_2+\lambda_3 {\bd q}_3{\bd q}_3,
 \eeq
 where $\{ {\bd q}_1, {\bd q}_2,{\bd q}_3\}$ is an orthonormal triad.  Three of the easy strains are pure shear: ${\bd q}_i{\bd q}_j +{\bd q}_j{\bd q}_i$, $i\ne j$ and the other two are
 $\lambda_3{\bd q}_2{\bd q}_2 -\lambda_2{\bd q}_3{\bd q}_3$ and $\lambda_1{\bd q}_2{\bd q}_2 -\lambda_2{\bd q}_1{\bd q}_1$.   Any other zero-energy strain is a linear combination of these.  
 Note that  there is no relation analogous to \rf{0-1} for strain in terms of stress because only the single ``component" ${\bd Q}: {\bd \varepsilon}$ is relevant, i.e., energetic. 

\begin{figure}[ht]     
\begin{center} \includegraphics[width=2.6in]{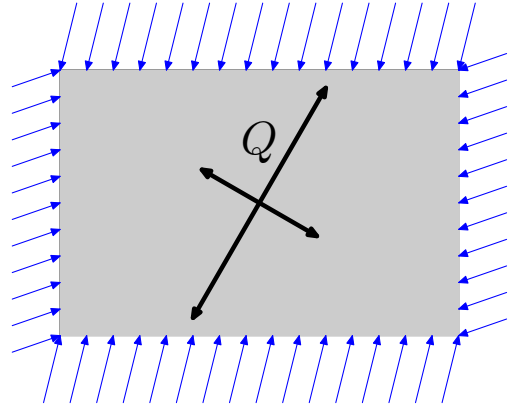} 
\end{center} 
\caption{A rectangular block of metafluid is  in static equilibrium under the action of surface tractions as shown.  The two orthogonal arrows inside the rectangle indicate the  principal directions of ${\bd Q}$ (30$^\circ$ from horizontal and vertical) and the  relative magnitude of its eigenvalues (2:1). The equispaced arrows faithfully represent the surface loads.   }
 \label{fig4}
\end{figure}

It is possible to write ${\bd \sigma}$ in the form \rf{0-4} where $p = - \kappa {\bd Q}: {\bd \varepsilon}$.
Under static load in the absence of body force  choose $\kappa$ such that $p=$constant, or equivalently, eq.$\rf{10}_2$. 
The relevant strain component is then 
${\bd Q}: {\bd \varepsilon} = - \kappa^{-1} p $  and the surface tractions supporting the body in equilibrium are 
${\bd t} = {\bd \sigma}\cdot {\bd n} = - p {\bd Q}\cdot {\bd n}$. Figure \ref{fig4} illustrates  the tractions required to maintain a block of metafluid in static equilibrium.   Note that the traction vectors act obliquely to the surface, implying that shear forces are necessary.  Furthermore, the tractions  are not of uniform magnitude.   These properties are to be compared with a normal acoustic fluid which can be maintained in static equilibrium by constant hydrostatic pressure.  

\subsubsection{$\div {\bd Q}=0$}
The \textit{gedanken} experiment of Fig.~\ref{fig4}  also implies  that ${\bd Q}$ has to be  divergence free.   Thus, imagine  a smoothly varying but  inhomogeneous metafluid 
${\bd C} = \kappa {\bd Q}\otimes {\bd Q}$ in static equilibrium under the traction ${\bd t} =  - p {\bd Q}\cdot {\bd n}$ for constant $p$.    The divergence theorem then implies $\div {\bd Q}=0$ everywhere in the interior.   This argument is a bit simplistic - but it provides the basis for a more rigorous proof\cite{Norris08b}.  Thus,  stress in the metafluid must be of the form $-f  \overline{\bd Q}$ where  
$ \overline{\bd Q}$ is a scalar multiple of  ${\bd Q}$.  Local equilibrium 
requires $\div f  \overline{\bd Q} =0$, or $\nabla \ln f = -  \overline{\bd Q}^{-1} \div  \overline{\bd Q}$. This can be integrated to find $f$ to within a constant.  Now define 
${\bd Q}= f \overline{\bd Q}$, and note that the tractions must be of the form 
${\bd t} =  - p {\bd Q}\cdot {\bd n}$ for constant $p$.   The normalized ${\bd Q}$ is divergence free. 

\subsubsection{Non-horizontal free surface}
Consider the same metafluid in equilibrium under a body force, e.g. gravity.   Assuming the inertia is isotropic (cf. the comments about inertia at zero frequency in \S \ref{sec6}),  
\beq{0-6}
\div {\bd \sigma} + \rho {\bd g} = 0. 
\eeq  
Use eq. $\rf{0-4}_2$ with $\div {\bd Q}=0$ and the invertibility of ${\bd Q}$, implies 
\beq{0+-}
\nabla p=    \rho {\bd Q}^{-1}{\bd g} . 
\eeq
For constant ${\bd Q}^{-1}\rho {\bd g}$  this can be integrated to give an explicit form for the pseudopressure,
\beq{0-7}
p = ({\bd x} - {\bd x}_0)\cdot   {\bd Q}^{-1} \rho {\bd g} , 
\eeq  
where ${\bd x}_0$ is any point lying on the surface of zero pressure.  Unlike normal fluids, the surface where $p=0$ does not have to be horizontal, see Fig.~\rf{fig5}. 
The pseudopressure increases in the direction of ${\bd g}$, as in normal fluids.  However, it is possible that $p$ varies in the plane ${\bd x}\cdot{\bd g} = 0 $.  For instance, the traction along the lower surface in   Fig.~\rf{fig5} decreases in magnitude from left to right.  

\begin{figure}[ht]  
\begin{center} \includegraphics[width=2.6in]{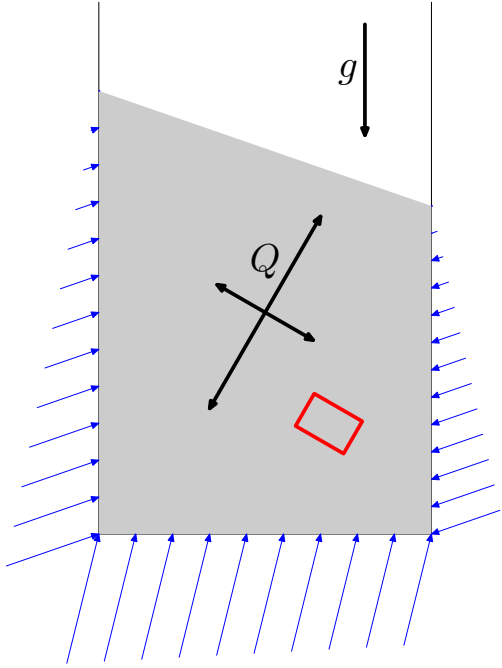} 
\end{center} 
\caption{The same metafluid of Fig.~\ref{fig4} with isotropic density is in equilibrium under gravity. The  upper surface  is traction free but non-horizontal, 
 an essential feature of metafluids.  For this particular metafluid the top surface makes an angle of 19.11$^\circ$ with the horizontal.  Also,  the tractions on the horizontal lower  boundary  are inhomogeneous although parallel. The small rectangle is discussed in Fig.~\ref{fig6}.}
 \label{fig5}   
\end{figure}   

\begin{figure}[ht]     
\begin{center} \includegraphics[width=2.6in]{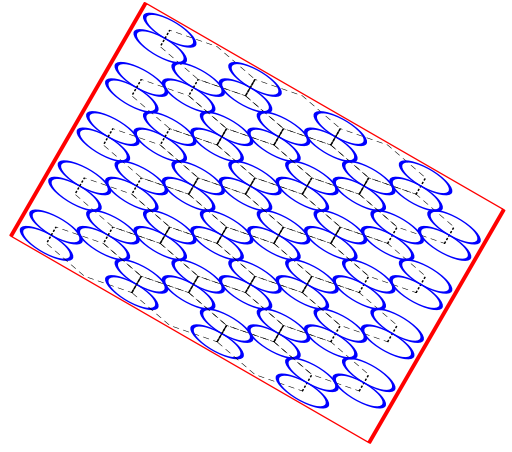} 
\end{center} 
\caption{This shows a possible microstructure for a microscopic rectangular region of the metafluid in Figs. \ref{fig4} and \ref{fig5}, see the latter. The sides of the rectangle are aligned with the principal axes of ${\bd Q}$.  The metafluid is a pentamode material comprising a regular array of small beads such that each is in  lubricated point contact with its three neighbors (2D).  The dashed lines indicate the directions of the forces acting between the small oval-shaped beads.  \rev{Although the structure as shown is unstable under shear, a realistic metafluid might contain some stabilizing mechanisms to enhance its rigidity.}  Details will be provided in a forthcoming paper.  }
 \label{fig6}
\end{figure}   

\section{Metafluids with isotropic density}\label{sec6}

\subsection{Necessary constraints on the finite deformation} 

The most practical case of interest is of course where the initial properties are those of a standard acoustic fluid with isotropic density and isotropic stress. The circumstances under which the mapped inertia ${\bd \rho}$ is also isotropic are now investigated. 
 Acoustic metafluids with isotropic inertia are an important subset since it can be argued that achieving anisotropic inertia could be more difficult than the anisotropic elasticity.  Indeed, the very concept of  anisotropic inertia is meaningless at zero frequency, unlike  anisotropic stiffness.  

Assuming ${\bd \rho}_0 = \rho_0 {\bd I}$ and ${\bd Q}_0= {\bd I}$ then the current density becomes, using eq. \rf{91b}  and  the fact that ${\bd Q}$ is symmetric, 
\beq{71}
{\bd \rho} = \rho_0 \Lambda {\bd Q}{\bd V}^{-2}{\bd Q}.
\eeq
If ${\bd \rho} = \rho {\bd I}$ then eq. \rf{71} implies ${\bd Q}$ must be of the form  
\beq{72}
{\bd Q} = \rho^{1/2}(\rho_0 \Lambda )^{-1/2}\,  {\bd V}.
\eeq
Thus, ${\bd Q}$ is proportional to the stretch tensor ${\bd V}$ and the coefficient of proportionality defines the current density.   

It is not in general possible to choose  ${\bd Q}$ in the form    
\beq{73}
{\bd Q} = \alpha   {\bd V}, 
\eeq
where $\alpha \ne 0$.   Certainly, ${\bd Q}$ of \rf{73} is symmetric and invertible but not necessarily divergence free.   The latter condition requires $\div (\alpha   {\bd V}) = 0 $ which in turn may be expressed $\nabla \ln \alpha = - {\bd V}^{-1} \div  {\bd V}$.  The necessary and sufficient condition that ${\bd V}^{-1} \div  {\bd V}$ is the gradient of a scalar function, and hence  $\alpha$ can be found which makes 
${\bd Q}$ of \rf{73} possible,  is that ${\bd V}$ satisfy 
\beq{733}
\curl {\bd V}^{-1} \div  {\bd V} = 0 .
\eeq
This condition is not very useful.  It does, however, indicate that the possibility of achieving isotropic $\bd \rho$ depends on the underlying finite deformation; there is a subset of general deformations that can yield isotropic inertia.  The deformation gradient ${\bd F} $ has nine independent elements, ${\bd V}$ has six, and ${\bd R}$ has three.  The condition \rf{733} is therefore a differential constraint  on six parameters.    We now demonstrate  an alternative statement of the condition $\div (\alpha   {\bd V}) = 0 $ in terms of the rotation ${\bd R}$.  This will turn out to be more useful, leading to  general forms of potential deformation gradients.

  Substituting ${\bd F} = \alpha^{-1}{\bd Q}{\bd R}$ into the identity $\rf{+2}_1$ and using \rf{22} implies that 
\beq{74}
Q_{ij}\, \big( \Lambda^{-1}  \alpha^{-1}R_{jK}\big)_{,i} = 0. 
\eeq
Using ${\bd Q} = \alpha   {\bd F}{\bd R}^t$ and the identity 
${\bd F}^t\nabla = \Nabla$ along 
with $\alpha \ne 0$, yields
\beq{75}
R_{jM} \big( \Lambda^{-1}  \alpha^{-1}R_{jK}\big)_{,M} = 0. 
\eeq
Then using the identity $(R_{jK}R_{jM}\big)_{,M} = 0$, eq. \rf{75} yields
\beq{76}
\beta_{,K} =  R_{jK} R_{jM,M}
\quad 
\leftrightarrow\quad
\Nabla \beta = {\bd R}^t \Div {\bd R}^t , 
\eeq
where $\beta = - \ln (\Lambda  \alpha)$. 

The necessary and sufficient condition that \rf{76} can be integrated to find $\beta$ is 
$\Nabla \wedge\Nabla \beta = 0 $, or using \rf{76} 
\beq{763}
\Curl {\bd R}^t \Div {\bd R}^t = 0. 
\eeq
The integrability condition \rf{763} is in general not satisfied by ${\bd R}$, except in trivial cases.  Norris \cite{Norris08b} noted that isotropic density can be obtained if ${\bd R}
= {\bd I}$.  This corresponds to $\beta =$constant, and it  can be realized more 
 generally if ${\bd R}$ is constant.  Hence, 
 \begin{lem}\label{lem1}
If the rotation ${\bd R}$ is constant then a normal acoustic fluid can be mapped to a unique metafluid with isotropic inertia: 
\beq{t7}
E_0 = \kappa_0 \big( \Div {\bd U}\big)^2 
 +  \rho_0  \dot{\bd U}\cdot\dot{\bd U}  \quad\text{in } \Omega,
\eeq  
is equivalent to  the current energy density 
\beq{78}
E = \lambda \big(  {\bd V}:\nabla {\bd u}\big)^2 
 + \rho   \dot{\bd u}\cdot\dot{\bd u}  \quad\text{in } \omega,
\eeq  
where 
\beq{90}
\lambda = \Lambda^{-1} \kappa_0 ,
\qquad
\rho  = \Lambda^{-1} \rho_0 .
\eeq
The  total mass of the deformed region $\omega$ is the same as the total mass contained in   $\Omega$.
\end{lem}
The parameter  $\lambda $ is used to distinguish it from $\kappa
 = \Lambda \kappa_0 $, because  in this case ${\bd Q} = \Lambda^{-1}{\bd V}$.  
Also, the displacement fields are related simply by $ {\bd u} = {\bd R}  {\bd U}$. 

As an example of a deformation satisfying Lemma \ref{lem1}: ${\bd x} = f\big(  {\bd X}\cdot
{\bd A}{\bd X}\big)\, {\bd A}{\bd X}$ for any constant positive definite symmetric ${\bd A}$.   This type of finite deformation includes the important cases of radially symmetric cloaks.  Thus, Norris \cite{Norris08b} showed that radially symmetric cloaks can be achieved using pentamode materials with isotropic inertia.   
 
 \subsection{General condition on the rotation}

The results so far indicate that isotropic inertia is achievable for transformation deformations with  constant rotation.  We would however like to understand the broader implications of 
eq. \rf{763}.  
The rotation can be expressed in Euler form 
\beq{81}
{\bd R} = \exp \big( {\theta \axt{\bd a} }\big), 
\eeq
where  $\theta$ is the angle of rotation, the unit vector  ${\bd a}$ is the rotation axis, and   
the axial tensor  $\axt ({\bd a})$ is a skew symmetric tensor defined by $\axt ({\bd a}) {\bd b} = {\bd a}\wedge {\bd b} $.  The vector $\theta{\bd a}$ encapsulates the three independent parameters in ${\bd R}$.  The integrability condition \rf{763} is 
 now replaced with a more explicit one in terms of $\theta ({\bd X})$ and ${\bd a} ({\bd X})$. 
It is shown in the Appendix that 
\beq{85}
{\bd R}^t\Div {\bd R}^t = 
  {\bd a}\wedge \Nabla\theta + {\bd Z}, 
  \eeq
where the vector ${\bd Z}$ follows from  eq. \rf{a2}.     In particular,    ${\bd Z}$ vanishes if the axis of rotation ${\bd a}$ is constant. 
In general, for arbitrary spatial dependence,  \rf{85} implies that the integrability condition 
\rf{763} is equivalent to the following constraint on the rotation parameters: 
\beq{86}
{\bd a}  \Nabla^2 \theta  
  -( {\bd a}\cdot \Nabla)\Nabla\theta
   +(\Nabla\theta \cdot \Nabla ) {\bd a}
  - ( \Nabla\cdot {\bd a})\Nabla\theta
  + \Curl {\bd Z}= 0.
  \eeq 
  In summary, 
   \begin{lem}\label{lem2}
If the rotation ${\bd R}$ satisfies \rf{763}, or equivalently, if 
   $\theta$ and ${\bd a}$  satisfy the condition  \rf{86}, then 
a normal acoustic fluid  can be mapped to a unique metafluid with isotropic inertia according to eqs.  \rf{t7} and \rf{78}
with 
\beq{977}
\lambda = \Lambda^{-1}  e^{-2\beta} \kappa_0 ,
\qquad
\rho  = \Lambda^{-1}  e^{-2\beta} \rho_0 ,
\eeq
 where the function  $\beta$ is defined by 
 eq. \rf{76}.
 \end{lem}

 \subsection{Simplification in  2D}
The integrability condition  \rf{763} simplifies for the important general configuration of two-dimensional spatial dependence.  In this case $ {\bd a} $ is constant, 
$ \theta  =  \hat{\theta}  ({\bd X}_\perp)$ where ${\bd X}_\perp \cdot {\bd a}=0$. 
 Equation \rf{86} then  reduces to 
\beq{87}
\Nabla^2 \hat{\theta} =0, 
\eeq

 \subsubsection{Example}
Consider finite deformations with  inhomogeneous rotation  
\beq{88}
\theta  = \theta_0 + \gamma X_1, 
\qquad  {\bd a} = {\bd e}_3,
\eeq
for constants $\theta_0 $ and $ \gamma $.  This satisfies \rf{87} and therefore eq. \rf{76} can be integrated.  
The metafluid in $\omega$ has isotropic density and pentamode stiffness given by Lemma \ref{lem2}, where $
 \beta = \gamma( X_2-X_{02})\big)$.
 The constants  $\theta_0$ and $X_{02}$ may   be set to zero, with no loss in generality. 

As an example of a deformation that has rotation of the form \rf{88}, 
consider deformation of the region $\Omega = [ -\frac{\pi}{2\gamma}, \frac{\pi}{2\gamma}]\times [0,L]\times \mathbb{R}$ according to 
\beq{802}
 \begin{pmatrix} 
x_1
\\
x_2
\\
x_3
\end{pmatrix}
= 
\begin{bmatrix} 
A_{11} & A_{12}& 0
\\
A_{12} & A_{22}& 0
\\
0 & 0 & \alpha \end{bmatrix}
 \begin{pmatrix} 
\gamma^{-1}\sin \gamma X_1\, e^{-\gamma X_2}
\\
\gamma^{-1}( 1- \cos \gamma X_1\, e^{-\gamma X_2})
\\
X_3
\end{pmatrix},  
\eeq
where $\alpha>0$ and the 2$\times$2 symmetric matrix ${\bd A}$ with elements $A_{ij}$ is positive definite.  The deformation gradient is ${\bd F}= {\bd V}(X_2){\bd R}(X_1)$ with 
${\bd V}  = {\bd A}e^{-\gamma X_2} + \alpha {\bd e}_3 {\bd e}_3$, and 
\beq{8040}
{\bd R}  = \begin{bmatrix} 
\cos \gamma X_1 & - \sin \gamma X_1& 0
\\
\sin \gamma X_1 & ~\cos \gamma X_1& 0
\\
0 & 0 & 1\end{bmatrix}.
\eeq
The mapped metafluid is defined by the energy density in $\omega$
\beq{806}
E = \frac1{\alpha \det {\bd A}}
\bigg[
\kappa_0 \big[ \big(  {\bd A}:\nabla {\bd u}\big)^2 e^{-2\gamma X_2} 
+ ( \alpha u_{3,3})^2\big] 
 + \rho_0   \dot{\bd u}\cdot\dot{\bd u} \bigg] .
 \nonumber
\eeq
In particular, it has isotropic inertia. 

This example is not directly applicable to modeling a complete acoustic cloak.  However, it opens up the possibility of patching together metafluids with different local properties, each with isotropic inertia so that the entire cloak has isotropic mass density. 

\section{Summary and conclusion}\label{sec7}

Whether it is the simple 1D acoustic mirage of Fig.~\ref{fig2} or a three-dimensional  acoustic cloak, we have seen how acoustic stealth can be achieved using the concept of domain transformation.  The fluid in the  transformed region  exactly replicates the acoustical properties of the original domain.
The most general class of material that describes both the mimic and the mimicked fluids is defined  as an acoustic metafluid.  A general procedure for mapping/transforming one acoustic metafluid to another  has been described in this paper. 

The results, particularly Theorem \ref{thm1} in \S \ref{sec4}, show that acoustic metafluids are characterized by as few as two parameters $(\rho ,\,   \sqrt{\kappa})$ and as many as twelve $({\bd \rho} ,\,  \sqrt{\kappa}{\bd Q})$.   This broad class of materials can be described as pentamode materials with anisotropic inertia.  It includes the restricted set of fluids with anisotropic inertia and isotropic stress $({\bd Q}= {\bd I})$.   

The arbitrary nature of the divergence free tensor ${\bd Q}$ adds an enormous amount of latitude to the stealth problem.  It may be selected in some circumstances to guarantee isotropic inertia in cloaking materials, examples of which are given elsewhere\cite{Norris08b}.  In this paper we have derived and described the most general conditions required for ${\bd \rho}$ to be isotropic.   The conditions have been phrased in terms of the rotation part of the deformation, ${\bd R}$.   If this is a constant then the cloaking metafluid is defined by Lemma \ref{lem1}.   Otherwise the condition is eq. \rf{763} with the  metafluid given by    Lemma \ref{lem2}.    The importance of being able to use metafluids with isotropic inertia should not be underestimated.  Apart from the fact that it resolves questions of infinite total \rev{effective} mass \cite{Norris08b} isotropic inertia removes frequency bandwidth issues that would be an intrinsic drawback in materials based on anisotropic inertia.  

This paper  also describes,  for the first time, some  of the unusual physical  features of acoustic metafluids.   Strange effects are to be expected in static equilibrium, as illustrated in Figs.~\ref{fig4} and \ref{fig5}.   These properties can be best understood through realization of acoustic metafluids, and a first step in that direction is provided by the type of microstructure depicted in Fig.~\ref{fig6}.  \rev{The macroscopic homogenized equations governing the microstructure are assumed in this paper to be those of normal elasticity.  It is also possible that the large contrasts  required in acoustic metafluids could be modeled with more sophisticated constitutive theories, such as non-local models  or theories involving higher order gradients.}  
 There is considerable progress to be made in the modleing, design and ultimate fabrication of acoustic metafluids. 

In addition to the degrees of freedom associated with the tensor $\bd Q$, the properties of metafluids depend upon the finite deformation through $\bd V$.  
Even in the simple example of the 1D mirage, one could arrive at the lower picture in Fig.~\ref{fig2} through different finite deformations.  This raises the question of how to best choose the nonunique deformation gradient ${\bd F}$.    The present results indicate some strategies for choosing $\bd F$ to ensure the cloak inertia has isotropic mass, and the cloaking properties are in effect determined by the elastic pentamodal material. Li and Pendry  \cite{Pendry08} consider other  optimal choices for the finite deformation.  
Combined with the enormous freedom afforded by the arbitrary nature of ${\bd Q}$, there are clearly many optimization strategies to be considered.

\section*{Acknowledgments}
Thanks to the Laboratoire de M\'ecanique Physique at the Universit\'e Bordeaux 1 for hosting the author, and in particular Dr. A. Shuvalov.  And to the reviewers.

\appendix
\section{Derivation of eq. \rf{85}}

Equation \rf{81} can be written in Euler-Rodrigues form \cite{Norris06}
\beq{83}
{\bd R} = {\bd I} + \sin\theta {\bd S} +(1-\cos\theta){\bd S}^2,
\eeq
where
\beq{82}
{\bd S} \equiv \axt ({\bd a}) = \begin{bmatrix}
0 & -a_3 & a_2 \\
 a_3 &0 & -a_1 \\
-a_2 & a_1 & 0
\end{bmatrix}. 
\eeq
Equation \rf{83} can be used, for instance, to find $\theta = \cos^{-1}[\frac12(1- \tr {\bd R})]$ and hence 
${\bd a}$ from ${\bd S} = ({\bd R}- {\bd R}^t)/(2\sin\theta)$.  
 For the sake of notational brevity, in  the remainder of the Appendix  $c=\cos \theta$  and  $s=\sin \theta$. 
 
Explicit differentiation of \rf{83} yields
\beq{a0}
R_{iJ,J} = (c S_{iJ}- sP_{iJ})\theta_{,J} + sS_{iJ,J}  - (1-c)P_{iJ,J} ,
\eeq
where ${\bd P} = - {\bd S}^2 =  {\bd I}- {\bd a} {\bd a}$. 
Noting that $S_{iJ,J} = - (\Nabla \wedge {\bd a})_i$ and 
$P_{iJ,J} = - a_i \Nabla\cdot {\bd a} - {\bd a}\cdot\Nabla a_i$, 
implies  
\bal{a1}
 \Div {\bd R}^t &= 
 c  {\bd a}\wedge \Nabla\theta - s \Nabla\wedge{\bd a}
 - s ({\bd I}- {\bd a} {\bd a})\cdot  \Nabla\theta
 \nonumber \\ & \quad
 +(1-c)\big[ {\bd a} (\Nabla\cdot{\bd a}) + ({\bd a}\cdot\Nabla){\bd a}\big].
\end{align}
Multiplying by ${\bd R}^t$ using \rf{83}, gives after some elimination and simplification, 
\beq{a20}
 {\bd R}^t\Div {\bd R}^t  = 
  {\bd a}\wedge \Nabla\theta  + {\bd Z}, 
  \eeq
where  
  \bal{a2}
{\bd Z}&=  -s[c{\bd I} + (1-c){\bd a} {\bd a}]\cdot  \Nabla\wedge{\bd a}
 +(1-c){\bd a} (\Nabla\cdot{\bd a})
  \nonumber \\
  &\quad 
   -(1-c)[{\bd I} - (1-c){\bd a} {\bd a}]\cdot({\bd a}\cdot\Nabla){\bd a}  
     \nonumber \\
  &\quad 
   - s(1-c){\bd a}\wedge ({\bd a}\cdot\Nabla){\bd a} .
\end{align}


\begin{thebibliography}{10}
\bibitem{endnote1}
The name pentamode is based on the defining property that the material supports
  five easy modes of infinitesimal strain. See \protect \S \ref {sec5} for
  details.
\bibitem{endnote2}
The bijective property of the mapping does not extend to acoustic cloaks, where
  there is a single point in $\Omega $ mapped to a surface in $\omega $, see
  Norris \cite {Norris08b} for details.
\end{thebibliography}

\end{document}